\RequirePackage{fix-cm}
\documentclass[smallextended]{svjour3}       
\smartqed  
\usepackage{graphicx}
\usepackage{epstopdf}
\begin{document}

\title{Temperature phase transition model for the DNA-CNTs-based nanotweezers
}


\author{Anh D. Phan, Lilia M. Woods and N. A. Viet}


\institute{Anh D. Phan and Lilia M. Woods \at
              Department of Physics, University of South Florida, Tampa, Florida 33620, USA \\
              Tel.: +1-813-974-8489\\
              Fax: +1-813-974-5813\\
              \email{anhphan@mail.usf.edu}           
           \and
           N. A. Viet \at
              Institute of Physics, 10 Daotan, Badinh, Hanoi, Vietnam
}

\date{Received: date / Accepted: date}

\maketitle

\begin{abstract}
DNA and Carbon nanotubes (CNTs) have unique physical, mechanical and electronic properties that make them revolutionary materials for advances in technology. In state-of-the-art applications, these physical properties can be exploited to design a type of bio-nanorobot. In this paper, we present the behaviors of DNA-based nanotweezers and show the capabilities of controlling the robotic device. The theoretical calculations are based on the Peyrard-Bishop model for DNA. Furthermore, the influence of the van der Waals force between two CNTs on the opening and closing of nanotweezers is studied in comparison with the stretching forces of DNA.
\end{abstract}

\section{Introduction}

In the past several years, researchers have made much progress in synthesizing new materials and developing fabrication techniques necessary for nanoscaled device production. This progress has been particularly important for applications  utilizing physical systems intended for biological and medical purposes.In this regard, biophysical devices at the nanoscale open up novel possibilities for diagnostic and therapeutic applications. 

DNA and carbon nanotubes (CNTs) are interesting and important systems in nanoscience. They have been the subject of many investigations in the past two decades\cite{1,2,3,4,23}. DNA is composed of two long polymer strands organized in a double helical structure, where each strand consists of repeating units (nucleotides)\cite{24}. CNTs are quasi-one dimensional cylindrically wrapped graphene sheets with properties uniquely defined by theregistry dependence of the wrapping given by a chirality index $(n,m)$ \cite{25}. Various applications of DNA/CNT complexes have been exploited with potential for biosensors \cite{6}, DNA transporters \cite{7}, and field effect transistors \cite{8}. The DNA/CNT is a composite with complicated structure with temperature dependent motion dynamics. Recently, using molecular dynamics simulations researchers have proposed molecular tweezers combining DNA and CNTs \cite{19} - a device with further technological and scientific potential.

A theoretical model of a geometrical soliton of DNA structure was constructed for the first time by Englander \cite{9} (E model). In this model, one of the strands of the DNA is represented as
a chain of pendula interacting with the another fixed similar strand. The E model explains the existence of DNA open state due to nonlinear excitations. In addition, the DNA structure and dyanmics has been modeled in terms of the Peyrard-Bishop (PB) model \cite{1}, which has been succssessful in explaining DNA denaturation transitions, pre-melting dynamics, and thermal transport. In the PB model, backbone of DNA is described as chains of particles with nearest neighboring potentials. However, the models ignore the helicoidal structure of the DNA molecule, the context of DNA flexibility, and the properties associated with it. 

CNTs are chemically inert and they interact with other materials via long-ranged dispersive forces, such as van der Waals (vdW) forces. The vdW interactions of  graphitic nanostructures can be described via pairwise interatomic Lennard-Jones (LJ) potentials \cite{11}. This approach relies on knowledge of the coupling Hamaker constants and it predicts the equilibrium separation correctly. The LJ potential has been applied to model mutual interaction between various CNTs as well as CNT based devices \cite{26,27}.

In this work we investigate the dynamics of hybrid DNA/CNT nanotweezers by employing the PB and vdW-LJ models. This dynamics of stretching in terms of its velocity and acceleartion due to environmental temperature changes is investigated. The critical temperature where a melting transition of the DNA/CNT takes place is presented. Comparisons between the strength of the involved forces showing the temperature-dependent motion is dominated by the stretching of the H bonds and bases, while the CNT vdW interaction is weaker. 

The rest of the paper is organized as follows: In Sec. II, the theoretical structure model, behavior and interactions of DNA-based nanotweezer are introduced. In Sec. III, numerical results are presented. The conclusions are given in Sec. IV.
\section{Model and mathematical background}
\begin{figure}[htp]
\includegraphics[width=9cm]{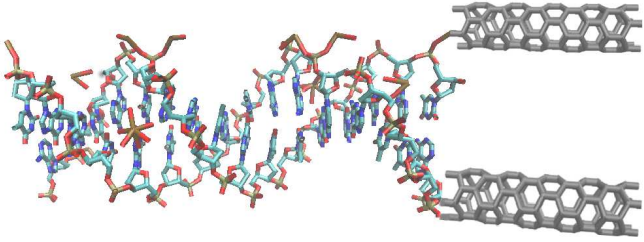}
\caption{\label{fig:0}(Color online) Schematics of the DNA/CNTs-based nanotweezer.}
\end{figure}

The proposed nanotweezer architecture is assembled by attaching the reactive ends of two single wall CNTs to the DNA strands as shown in Fig.~\ref{fig:0}. The rest of the end C bonds are saturated via H atoms. The size of this hybrid is quite large, approximately thousands atoms, thus full quantum mechanical atomistic treatment is not possible. The PB model is relatively simple \cite{1}, which describes the DNA two strands as a coupled pendulum system. 
\subsection{Model of DNA dynamics}
According to the PB model \cite{1}, the DNA double strand is modeled by two parallel chains of nucleotides via nearest-neighbor harmonic oscillator interactions. The potential for the Hydrogen bonds is also included. The relevant Hamiltonian is given as follows \cite{1,2}
\begin{eqnarray}
H &=& \sum_{n=1}^{N}\left[\frac{1}{2}m\left(\dot{u}_{n}^2+\dot{v}_{n}^2\right)+\frac{1}{2}k\left(u_{n}-u_{n-1}\right)^2+\right.
\nonumber \\
& & \left.\frac{1}{2}k\left(v_{n}-v_{n-1}\right)^2 + V\left(u_{n}-v_{n}\right)\right],
\label{eq:1}
\end{eqnarray}
where $u_{n}$ and $v_{n}$ are the nucleotide displacements from equilibrium along the direction of the hydrogen bonds for each strand. $m$ is the mass of each nucleotide (taken to be the same for each unit), while $k$ is harmonic oscillator coupling constant of the nearest-neighbor longitudinal interaction along each strand in units of $eV/\AA^{2}$. The potential for the Hydrogen bonds between the two strands is modeled via a Morse potential $V(r)=D[e^{-\alpha r} - 1]^2$. Here, $D$ is the dissociation energy and $\alpha$ is a parameter. It is important to note that the Morse potential represents the hydrogen bonds between complementary bases, the repulsive interactions of the phosphate, and the influence of the solvent environment. 

The dynamics of the system described by Eq.(\ref{eq:1}) is conveniently described using a set of new variables $x_{n} = (u_{n}+v_{n})/\sqrt{2}$ and $y_{n} = (u_{n}-v_{n})/\sqrt{2}$, representing the in-phase and out-of-phase motion of the two strands, respectively. Using this separation of variables, the Hamiltonian is decoupled. An important point is that $y_{n}$ represents the relative displacements between two nucleoid at the site $n$ in different strands. It reflects the stretching of DNA. Here we consider the out-of-phase displacements stretch of the hydrogen bonds given by $H_y$ 

\begin{eqnarray}
H_{y} = \sum_{n=1}^{N}\left[\frac{1}{2}m\dot{y}_{n}^2+\frac{1}{2}k\left(y_{n}-y_{n-1}\right)^2 + V\left(2y_{n}\right)\right].
\label{eq:2}
\end{eqnarray}

In the case of large number of nucleotides $N \rightarrow \infty$ and $H$ is independent on the particular site $n$. Perfroming statistical averaging in the canonical ensemble, the Schrodinger equation of a single mode $y$  using $H_y$ is given by \cite{1,15,16}
\begin{eqnarray}
\left(-\frac{1}{2\beta^{2}k}\frac{\partial^{2}}{\partial y^{2}} + V(2y)\right)\varphi(y) = \varepsilon\varphi(y),
\label{eq:3}
\end{eqnarray}
where, $\beta = 1/k_{B}T$, and $k_{B}$ is the Boltzmann constant. The exact solution for eigenenergies is given \cite{26}
\begin{eqnarray}
\varepsilon_{n}=\frac{1}{2\beta}\ln\left(\frac{\beta k}{2\pi}\right)+\frac{2\alpha}{\beta}\sqrt{\frac{D}{k}}\left(n+\frac{1}{2}\right)-\frac{\alpha^2}{\beta^2k}\left(n+\frac{1}{2}\right)^2.
\label{eq:31}
\end{eqnarray}

Eq.(\ref{eq:31}) has a discrete energy spectrum when $d = (\beta/\alpha)\sqrt{kD} > 1/2$. This allows one to obatin a critical temperature $T_{c} = 2\sqrt{kD}/(\alpha k_{B}) $, which is considered as the melting temperature of DNA. The DNA states are continuous for $T > T_{c}$ and discrete for $T < T_{c}$. For the parameters of DNA, when we consider $T > 200$ $K$, only the value of $n = 0$ is taken into account. There is no excitation state for DNA in our considerations.
 
From this, the ground state eigenfunction and eigenenergy in the thermodynamics limit of a large system is obtained as \cite{1,2}
\begin{eqnarray}
\varphi_{0}(y)=\sqrt{\sqrt{2}\alpha}\frac{(2d)^{d-1/2}}{\sqrt{\Gamma(2d-1)}}e^{-de^{-\sqrt{2}\alpha y}}e^{-(d-1/2)\sqrt{2}\alpha y},
\label{eq:4}
\end{eqnarray}
\begin{eqnarray}
\varepsilon_{0}=\frac{1}{2\beta}\ln\left(\frac{\beta k}{2\pi}\right)+\frac{\alpha}{\beta}\sqrt{\frac{D}{k}}-\frac{\alpha^2}{4\beta^2k}.
\end{eqnarray}

In addition, the system described via Eq.(\ref{eq:3}) can be represented as a quasiparticle with a tempereture dependent effective mass $m^{*} = \hbar^{2}\beta^{2}k$.  At room temperature, the value of the effective mass is approximately $22.87$ $m_{0}$, here $m_{0}$ is the rest mass of electron. The average stretching of the hydrogen bonds can also be calculated via $\left\langle y\right\rangle = \int\varphi_{0}^{2}(y)ydy$ \cite{1,2}. 

The stretching force is determined via the expression
\begin{eqnarray}
F_{s} = -\frac{\partial V(\left\langle y\right\rangle)}{\partial \left\langle y\right\rangle}.
\label{eq:8}
\end{eqnarray}

\begin{figure}[htp]
\centering
\includegraphics[width=9cm]{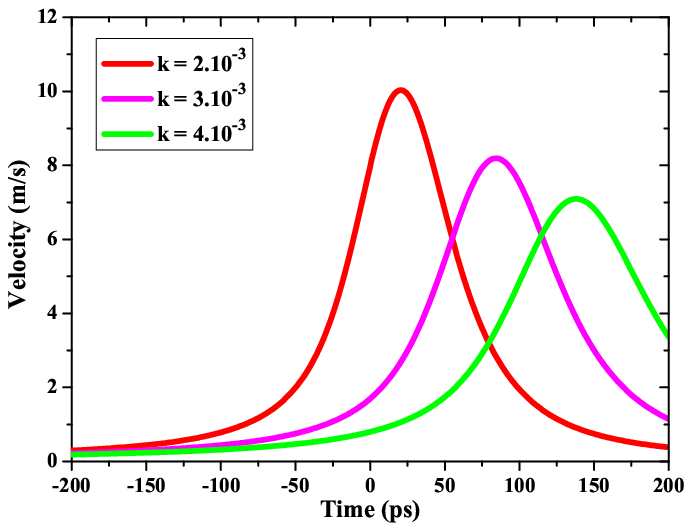}
\hspace{1in}%
\includegraphics[width=9cm]{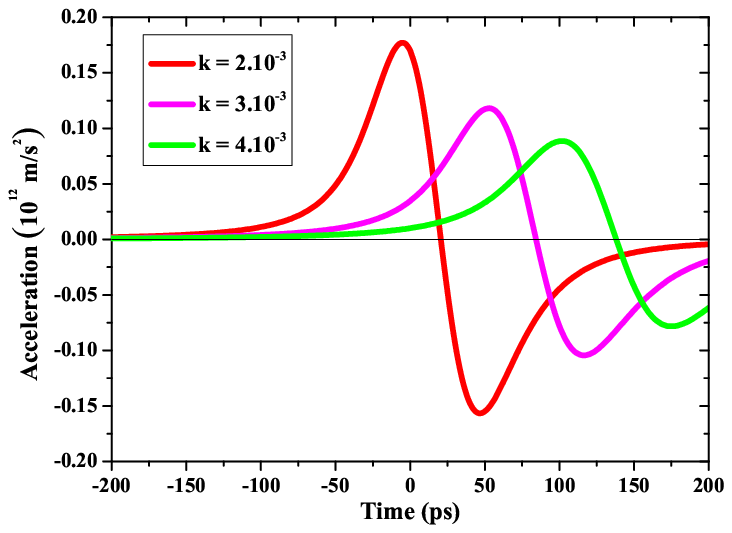}
\caption{\label{fig:2}(Color online)  (Color online) The time-dependent velocity and acceleration of the opening.}
\end{figure}

To investigate thermal properties of DNA, we heated up and cooled down temperature of the bio-systems flollowing an expression $T = 1.14t + 300$ (K) \cite{19}. Here $T$ (K) is the environment temperature, $t$ (ps) is time. Basing on the average stretching $\left\langle y\right\rangle$ of the coupling constants pointed out above, the velocity $v = d\left\langle y\right\rangle/dt$ and acceleration $a = d^{2}\left\langle y\right\rangle/dt^{2}$ of the opening of the nanotweezers obtained by taking the first and second derivative of the stretching with respect to time, respectively, are presented in Fig.~\ref{fig:2} 

For $k = 2.10^{-3}$ eV/$\AA^2$, the velocity of the opening increases and reaches to the maximum with the value of $10.34$ m/s at around $t = 20$ s. After that, the velocity drops significantly to zero. It refers that the temperature corresponding to the peak is $322.5$ K. On the other hand, initially, the value of acceleration is positive and rises to the maximum value $0.187\times10^{12}$ $m/s^2$ at $-5.2$ s or $294$ K before declining gradually to the negative side, crossing the time axis at $19.4$ s or $322.5$ K, touching the bottom $-0.164\times10^{12}$ $m/s^2$ at around $45$ s and continuing to approach to $0$. It can be easily explained due to the fact that below $322.5$ K, the stretching velocity climbs significantly, so the acceleration is positive. Zero acceleration, of course, is at the relevant bending point of the opening velocity. Above $322.5$ K, the unzipping velocity declines notably, and is nearly unchanged. Therefore, the acceleration has the negative values and goes to zero. 

In the same way, for other values of $k = 3.10^{-3}$  eV/$\AA^2$ and $k = 4.10^{-3}$ eV/$\AA^2$, the zero acceleration takes place at $88$ $s$ and $138$ $s$, respectively. It means that the melting temperatures corresponds to $401$ $K$ for $k = 3.10^{-3}$ eV/$\AA^2$ and $456$ $K$ for $k = 4.10^{-3}$  eV/$\AA^2$. As a result, there is a possibility to obtain the melting temperature by observing the velocity of stretching.

\subsection{CNT van der Waals interaction}
The vdW interaction between the CNT parts of the DNA nanotweezers is described via the Lennard-Jones (LJ) approximation. This approach is widely used in calculating disperssive interactions between graphitic nanostructures because of its relative simplicity and satisfactory results in determining their equilibrium configurations \cite{11}. The LJ potential is essentially a pairwise apprximation, and for extended systems, one typically perfroms integration over the volumes of the interacting objects. For CNTs, the integration is over the surfaces of hollow cylinders with radii corresponding to the radii of the nanotubes. The LJ-vdW potential per unit length for two parallel CNTs with radii $R_1$ and $R_2$ is given by \cite{10}
\begin{figure}[htp]
\includegraphics[width=9cm]{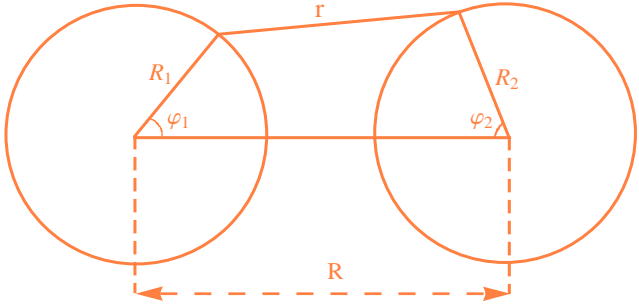}
\caption{\label{fig:00}(Color online) Sketch of van der Waals interaction between two CNTs.}
\end{figure}

\begin{eqnarray}
V_{vdW}=\sigma^2\int\int\left(-\frac{A}{\rho^6}+\frac{B}{\rho^{12}}\right)dS_{1}dS_{2},
\end{eqnarray}
where $A$ and $B$ are the Hamaker constants corresponding to the attractive and repulsive contributions, respectively.   For graphitic systems, one typically takes the values for graphite $A = 15.2$ eV$\AA^6$ and $B = 24\times10^3$ eV$\AA^{12}$ \cite{11}. $\sigma = 4/\sqrt{3}a^2$ is the mean surface density of Carbon atoms with $a = 2.49$ $\AA$ being the lattice constant. Also, the distance between the CNT surfaces is $\rho$. Perfroming the integration over the length of the two CNTs with radii $R_{1}$ and $R_{2}$, the LJ-vdW interaction can be written as \cite{10}: 
\begin{eqnarray}
V_{vdW} &=& -\frac{3\pi A\sigma^{2}R_{1}R_{2}}{8}\int_{0}^{2\pi}\int_{0}^{2\pi}\frac{1}{r^5}d\varphi_{1}d\varphi_{2}
\nonumber \\
& & + \frac{63\pi B\sigma^{2}R_{1}R_{2}}{256}\int_{0}^{2\pi}\int_{0}^{2\pi}\frac{1}{r^{11}}d\varphi_{1}d\varphi_{2},
\label{eq:7}
\end{eqnarray}
where the in-plane distance between two surface elements is defined as $r^2 = (R-R_{1}\cos\varphi_{1}+R_{2}\cos\varphi_{2})^2+(R_{1}\sin\varphi_{1}-R_{2}\sin\varphi_{2})^2$. The definitions of $R_{1}$, $R_{2}$, $\varphi_{1}$, $\varphi_{2}$, and $r$ are sketched in Fig.~\ref{fig:00}.

Then, applying the first derivative with respect to $R$, we obtain the van der Waals interaction force per unit length
\begin{eqnarray}
F_{vdW}(R) = -\frac{\partial V_{vdW}}{\partial R}.
\end{eqnarray}

\section{Numerical results and discussions}
As a prototype, we take that both CNTs are identical with the chiral vector $(5,0)$ and $(6,0)$, and lengths $L_{1} = L_{2} = 5$ $nm$. The total Hamiltonian for the system is composed of two term, that account for the stretching and van der Waals interaction - $H=H_y+V_{vdW}$. Because of the relatively weak vdW force between the tubes, $V_{vdW}$ is treated as a perturbation compared to $H_y$. The parameters of DNA are $D = 0.33$ $eV$ and $\alpha = 18$ $nm^{-1}$. It is important to note that $\varphi_{0}(y)$ and $\varepsilon_{0}$ in the previous section is the wave function and energy of the ground state of DNA without the presence of CNTs.   

\begin{figure}[htp]
\includegraphics[width=9cm]{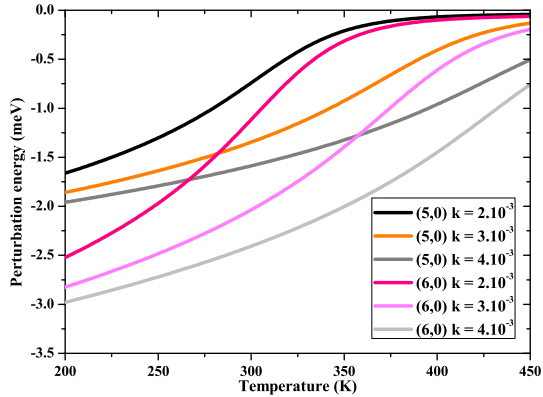}
\caption{\label{fig:1}(Color online) The first-order energy is caused by the van der Waals interactions between two CNTs.}
\end{figure}

\begin{figure}[htp]
\includegraphics[width=9cm]{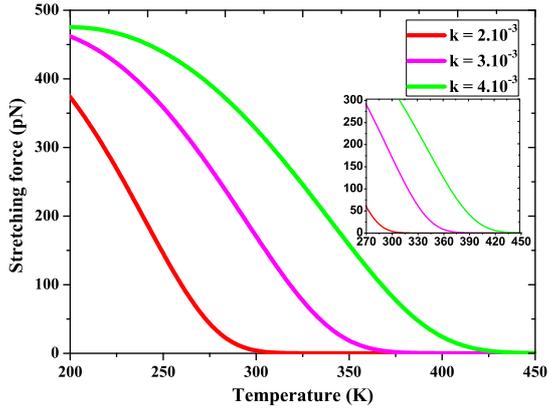}
\caption{\label{fig:4}(Color online) The unzipping force as a function of temperature .}
\end{figure}

In Fig.~\ref{fig:1} and Fig.~\ref{fig:4}, we show results for the CNT vdW perturbative force correction as a function of tempertaure and the stretching force. Fig.~\ref{fig:4} indicates that $F_s$ decreases as $T$ increases. The stretching force goes to zero at the critical temperature since the properties of DNA change when $T$ reaches to $T_{c}$. 

Obviously, the wave function is temperature-dependent, so the energy and energy shift are functions of temperature. The value of $\varepsilon_{0}$ for three values of $k$ at this range of temperature varies from $220$ meV to $280$ meV. It means that the influence of the van der Waals interaction on the wave function and the energy in the ground state is minor. We can calculate separately the interactions of DNA and CNTs. An additional point is that the larger temperature is, the smaller the first-order pertubation of energy is. A simple reason for this problem is that when temperature increases, two DNA strands are opened \cite{1} and it leads to a rapid growth of distance between two CNTs. 

It is remarkable that we have studied the van der Waals interaction and the pertubation energy between two parallel CNTs. This configuration also is used in order to calculate all of the van der Waals interactions below. Nevertheless, in actual cases, we have two crossed CNTs. The dispersion interaction in real biosystems is weaker than that in the parallel state. Therefore, we can utilize the wave function $\varphi_{0}(y)$ in the following calculations without addional terms due to the perturbation theory.

It is clearly seen in Fig.~\ref{fig:4}, at the critical temperature $T_{c}$, the stretching force vanishes because two strands of DNA are broken for $T > T_{c}$. The opening force of DNA is very large at low temperature. The smaller the temperature is, the smaller distance between two strands is. This force decreases when increasing temperature since the separation distance is larger and larger.

These results have aggrement about the range of magnitude force with experimental data and previous calculations \cite{21,22}. The increase of $k$ causes to the growth of stretching force due to the fact that the binding of DNA rises.

\begin{figure}[htp]
\includegraphics[width=9cm]{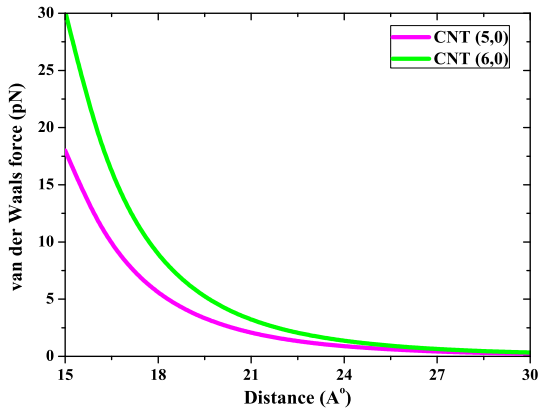}
\caption{\label{fig:5}(Color online) The van der Waals forces between two parallel CNTs (5,0) and (6,0) as a function of the separation distance between two centers of CNTs.}
\end{figure}

Lets consider the interaction between two CNTs attached in the ends of DNA. There are several types of DNA existing in nature such as B-DNA and Z-DNA. Since the diameter of DNA is approximately $2.37$ nm for B-DNA and $1.84$ nm for Z-DNA. We assume that the initial distance between two centers of CNTs is $1.5$ nm. It is important to note that the van der Waals force is attractive at this range of distance and the sign of this force should be minus. The magnitude of van der Waals interaction between two CNTs is presented in Fig.~\ref{fig:5}.

For $k = 2.10^{-3}$ eV/$\AA^2$, if $T < 277$ $K$, the stretching force is much larger than the van der Waals force of CNTs (5,0) and (6,0) at the initial state. Therefore, it is easy to control the opening and closing of DNA by cooling down or heating up. At low temperature, the contribution of the dispersion force in the movement of DNA strands is minor. However, it can rise to significant role when $T > 277$ $K$. We can do the same way with $k = 3.10^{-3}$ eV/$\AA^2$ and  $k = 4.10^{-3}$ eV/$\AA^2$.

\begin{figure}[htp]
\includegraphics[width=9cm]{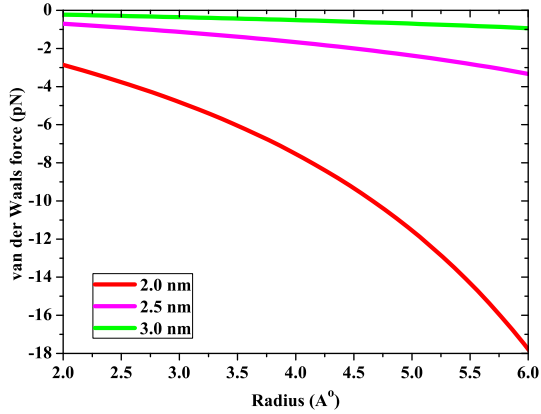}
\caption{\label{fig:6}(Color online) The van der Waals forces between CNT (5,0) and another CNT.}
\end{figure}

Figure \ref{fig:6} shows the forces between CNT (5,0) and different CNTs at the certain distances. In order to control the opening and closing of nanotweezers, the van der Waals force is weaker than the stretching forces. It is difficult to operate the movement of nanotweezers if two CNTs have large radii. 

When we heat up the biosystem, two ends of DNA are separated by the stretching force. At the larger temperature, the unzipping force is much larger than the van der Waals interactions, the nanotweezers are opened. The obtained results agree with the previous simulation study \cite{19}. Therefore, in our nanorobots, the movements of CNTs can be controlled by changing temperature. In addition, the van der Waals interaction between two cylinders is proportional to the length of tubes. If we want to have the smaller van der Waals interaction, it is possible to choose the length 1 nm or 2 nm. Another point is that long CNTs are bent because of the van der Waals interaction. As a consequence, the length of tubes should not be large in designing the bio-nanorobots.

\section{Conclusions} 
The use of intelligence, sensing and actuation nanodevices in surgery, medical treatments and materials science is a reality which has become a hot topic in the biomedical industry and research in recent years. Bio-nanorobots provide further advance not only in the nanotechnology, but also efficient approaches for disease treatment. Our studies showed the behavior and architecture of the bio-nanotweezers. The temperature dependence of the opening displacements of tweezers is presented and gives researchers some principles to understand the operation of DNA-based molecular machines and devices. In addition, the velocity and acceleration of the opening and closing tweezers as a function of time are speculated. The theoretical calculations are easy to understand and agree qualitatively with the previous works. Further research on these systems can considerably extend interdisciplinary implications for the technology.
\begin{acknowledgements}
We thank Professor M. Peyrard for helpful discussions and comments. We gratefully acknowledge support through the Department of Energy under Contract No. DE-FG02-06ER46297. The work was partly funded by the Nafosted Grant No. 103.06-2011.51.
\end{acknowledgements}

\end{document}